\documentclass{article}
\usepackage[numbers,sort&compress]{natbib}
\usepackage{spconf,amsmath,epsfig}
\usepackage{booktabs}
\usepackage{multirow}
\usepackage{subfigure}
\usepackage{graphicx}
\usepackage{float}
\usepackage{amsmath, bm}
\usepackage{balance}
\usepackage{amsfonts}
\usepackage{tabularx}
\usepackage{url}

\newcommand{\tabincell}[2]{\begin{tabular}{@{}#1@{}}#2\end{tabular}}

\title{Automatic Depression Detection: An Emotional Audio-Textual Corpus and a GRU/BiLSTM-based Model}
\name{Ying Shen, Huiyu Yang, Lin Lin$^{*}$ \thanks{$*$ Corresponding author.}\thanks{This work was supported in part by the National Natural Science Foundation of China under Grants 61972285, in part by the Natural Science Foundation of Shanghai under Grant 19ZR1461300, in part by the Shanghai Science and Technology Innovation Plan under Grant 20510760400, in part by the Shanghai Municipal Science and Technology Major Project under Grant 2021SHZDZX0100, and in part by the Fundamental Research Funds for the Central Universities.}}
\address{School of Software Engineering, Tongji University, P.R.China \\
\url{{yingshen, 2031552, 1931542}@tongji.edu.cn}
}

\begin{document}
%
\ninept
\maketitle

\begin{abstract}
Depression is a global mental health problem, the worst case of which can lead to suicide.
An automatic depression detection system provides great help in facilitating depression self-assessment and improving diagnostic accuracy.
In this work, we propose a novel depression detection approach utilizing speech characteristics and linguistic contents from participants' interviews.
In addition, we establish an Emotional Audio-Textual Depression Corpus (EATD-Corpus) which contains audios and extracted transcripts of responses from depressed and non-depressed volunteers.
To the best of our knowledge, EATD-Corpus is the first and only public depression dataset that contains audio and text data in Chinese.
Evaluated on two depression datasets, the proposed method achieves the state-of-the-art performances.
The outperforming results demonstrate the effectiveness and generalization ability of the proposed method.
The source code and EATD-Corpus are available at \url{https://github.com/speechandlanguageprocessing/ICASSP2022-Depression}.
\end{abstract}
\section{Introduction}
\label{sec:intro}

Depression is a common mental disorder, the three main symptoms of which are persistent low mood, loss of interest and lack of energy \cite{peveler02,wu20}.
In the worst case, depression can lead to suicide.
According to World Health Organization reports, about 264 million people are suffering from depression worldwide \cite{WHO}.
However, the treatment rate of depressed people remains very low in the whole world \cite{kessler12}. 
There are mainly two factors accounting for the low treatment rate.
Firstly, traditional treatments for depression are time-consuming, costly and sometimes ineffective \cite{Craft1998}.
The cost of diagnosis and treatment can be a heavy burden for individuals with financial difficulties, and thus makes them reluctant to seek help from physicians.
Secondly, during the clinical interviews of depression diagnosis, patients may hide their real mental states in fear of prejudice or discriminatory behaviors towards the depressed people \cite{yokoya18,Haque2019}.
In such cases, the clinician is unable to make a correct diagnosis.
The aforementioned factors have necessitated the automatic depression detection system, which can help individuals assess their depressive states privately as well as increase their willingness to consult the psychologists.
Furthermore, such a system would be of great help to psychologists in depression diagnosis when patients hide their real mental states.


\section{Related Work and Our Contributions}
\label{sec:relatedwork}

\noindent \textbf{Automatic depression detection.}
Early studies of automatic depression detection were dedicated to extracting effective features from questions that were highly correlated with depression.
Sun {\itshape et al.} \cite{sun2017} conducted content analysis to the text transcripts of clinical interviews and manually selected questions related to certain topics (e.g. Sleeping quality or recent feelings).
Based on the text features extracted from the selected questions, they used Random Forest to detect depression tendency.
Similarly, Yang {\itshape et al.} \cite{yang2016} also manually selected depression related questions after analyzing interview transcripts.
They constructed a decision tree with the selected questions to predict the participants' depression states.
Gong and Poellabauer \cite{gong2017} performed topic modeling to split the interviews into topic-related segments, from which audio, video, and semantic features are extracted.
They employed a feature selection algorithm to maintain the most discriminating features.
Williamson {\itshape et al.} \cite{williamson2016} constructed semantic context indicators related to factors such as depression diagnosis, medical/psychological therapy or negative feelings.
Utilizing Gaussian Staircase Model, they achieved a good performance in depression detection.

Inspired by the emerging deep learning techniques, integrating multi-modal features through deep learning models is particularly promising for depression detection.
Yang {\itshape et al.} \cite{yang2017} presented a depression detection model based on deep Convolution Neural Network (CNN).
They additionally designed a set of audio and video descriptors to train their model.
Tuka {\itshape et al.} \cite{Al2018} proposed a Long Short-Term Memory (LSTM) network to assess depression tendency.
They calculated Pearson Coefficients to select audio features and text features that were strongly related to depression severity.
With the combination of CNN and LSTM, Ma {\itshape et al.} \cite{ma2016} encoded the depressive audio characteristics to predict the presence of depression.
Haque {\itshape et al.} \cite{Haque2019} proposed a causal CNN model which summarized acoustic, visual and linguistic features into embeddings which were then used to predict depressive states.

\noindent \textbf{Our motivations and contributions.}
In the field of automatic depression detection, several limitations exist in current research. 
First of all, some methods rely heavily on manually selected questions which requires psychologists' expertise involved. 
Besides, all these preset questions have to be answered during the interview, otherwise the analysis may fail. 
How to improve detection performance without preset questions remains a challenging task.
In addition, publicly available depression datasets are scarce due to ethic issues. 
In this work, we make efforts to overcome the aforementioned drawbacks:

(1) To facilitate study of depression detection, we first establish EATD-Corpus, a publicly available Chinese depression dataset, which comprises audios and text transcripts extracted from the interviews of 162 volunteers.

(2) We then propose a novel method for automatic depression detection.
In this method, a Gate Recurrent Unit (GRU) model and a Bidirectional Long Short-Term Memory (BiLSTM) model with an attention layer are utilized to summarize representations from audio and text features. 
In addition, a multi-modal fusion network 
integrates the summarized 
features to detection depression.

\section{EATD-Corpus}
\label{EATD-Corpus}

The depression datasets are quite scarce \cite{yang2012,Gratch2014,Valstar2013,alghowinem2013,Scherer2013}.
To the best of our knowledge, there are only two publicly available datasets referring to depression detection.
The first one is DAIC-WoZ which contains recordings and transcripts of 142 American participants who were clinically interviewed by a computer agent \cite{Gratch2014}. 
The second one is AViD-Corpus \cite{valstar2014} which also contains audios and videos of German participants answering a set of queries or reciting fables.
However, the transcripts are not provided by the authors. 

In this work, we release a new Chinese depression dataset, namely EATD-Corpus, to facilitate the research in depression detection. 
EATD-Corpus consists of audios and text transcripts extracted from the interviews of 162 student volunteers recruited from Tongji University.
All the volunteers have signed informed consents and guarantee the authenticity of all the information provided.
Each volunteer is required to answer three randomly selected questions and complete an SDS questionnaire.
The SDS questionnaire consists of 20 items which rate the four common characteristics of depression: the pervasive effect, the physiological equivalents, other disturbances, and psychomotor activities \cite{zung1965}.
SDS is a commonly used questionnaire for psychologists to screen depressed individuals in practise.
A raw SDS score can be summarized from the questionnaire.
For Chinese people, an index SDS score (i.e. raw SDS score$\times$1.25) greater than or equal to 53 implies that he/she is in depression \cite{wang09}.
According to the criterion, there are 30 depressed volunteers and 132 non-depressed volunteers in EATD-Corpus.
The overall duration of response audios in the dataset is about 2.26 hours.

%
The process of constructing EATD-Corpus consists of two steps: data collection and data preprocessing.


\textbf{Data collection.}
An APP, through which a virtual interviewer will ask the interviewee three questions, is developed to conduct the interview and to collect audio responses.
The interviewees can record their responses and upload the response audios online.
Besides, each volunteer is required to complete an SDS questionnaire, the score of which indicates the depression severity.
Currently, 162 volunteers have successfully finished online interviews. 
Based on their SDS scores, 30 volunteers are regarded in depression and the other 132 volunteers are non-depressive.


\textbf{Data preprocessing.}
Several preprocessing operations have been performed on the collected audios.
First, mute audios, audios less than 1 second, 
and the silent segments at the beginning and the end of each recording are removed.
Then the background noises are eliminated using RNNoise \cite{Valin2018} with default parameters.
After that, Kaldi \cite{Povey2014} is used to extract transcripts from the audios.
In the end, all the transcripts were manually checked and corrected.

\section{A Multi-modal Depression Detection Method}
\label{multi-modal_method}

\begin{figure*}
\centering
\includegraphics[width=160mm]{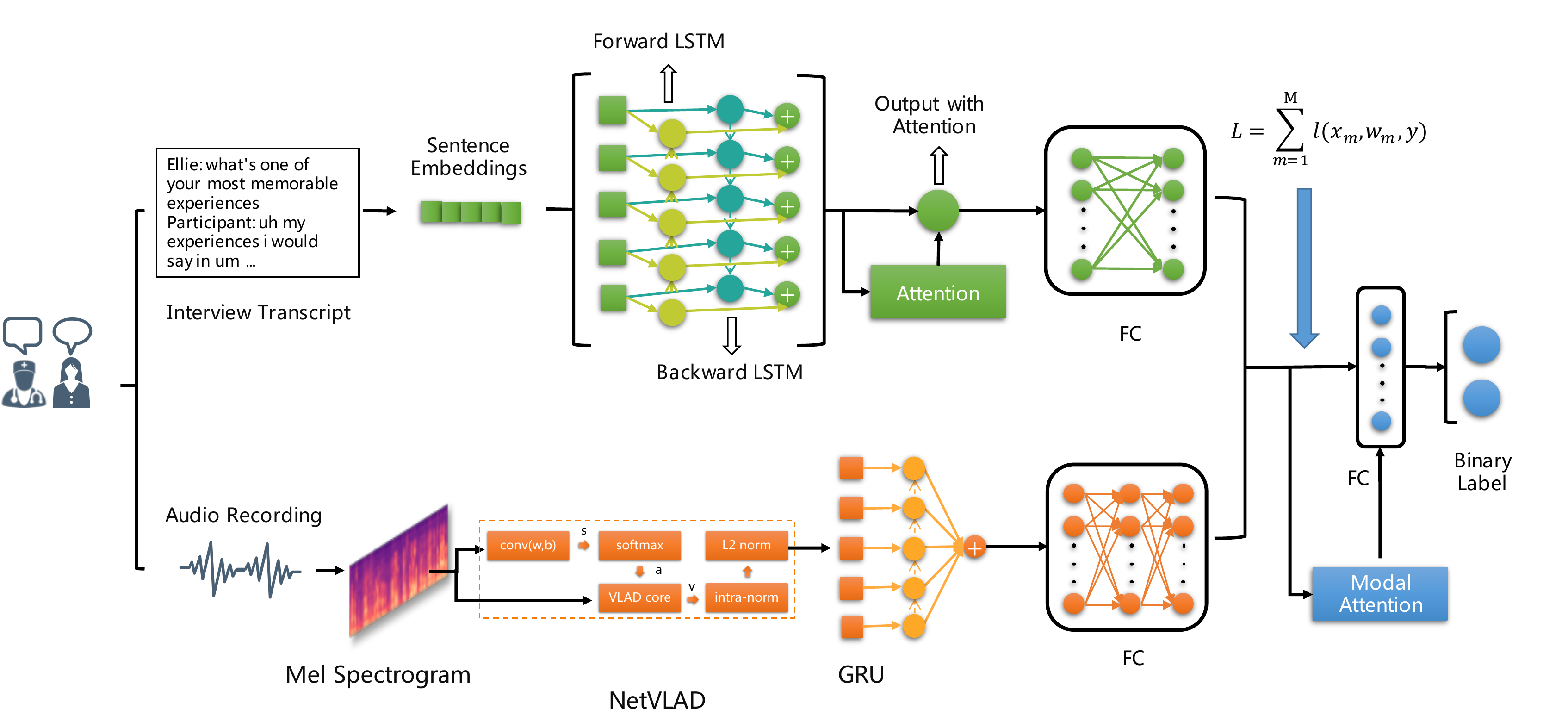}
\caption{Structure of the proposed model. 
Text features are trained using a BiLSTM model with an attention layer. 
Audio features are trained with a GRU model. 
Outputs from both BiLSTM and GRU models are concatenated. Modal attention is trained additionally to assign weights to outputs from the two models. The dot products of modal attention and concatenated outputs are fed into a fully connected network (FC) to generate binary labels.}
  \label{structure}
\vspace{-4mm}
\end{figure*}

In addition, we propose an efficient method for automatic depression detection. As shown in Fig.\ref{structure}, the proposed approach consists of a GRU model and a BiLSTM model with an attention layer.
The two models summarize audio and text representations, which are then concatenated and passed to a one-layer fully connected (FC) network.
Modal attention is a trained weight vector predicting the importance of two modalities.
The FC network outputs a binary label indicating the presence of depression.

\vspace{-2mm}
\subsection{Features}
In our method, text and audio features are used to prediction depression state.
Text features are extracted by projecting transcript sentences into high-dimensional sentence embeddings using ELMo \cite{matthew2018}.
For audio features, Mel spectrograms are extracted from the audios.
However, the sizes of the extracted Mel spectrograms vary greatly because the lengths of the audios range from 2 seconds to 1 minute.
Therefore, NetVLAD \cite{aran2018} is further adopted to generate audio embeddings of the
same length from Mel spectrograms. 

\subsection{BiLSTM with Attention Layer}

To extract text features, BiLSTM with an attention layer is adopted to emphasize which sentence contributes most in depression detection. 

\begin{equation}
\label{atten}
\begin{aligned}
  \mathbb{O} &= \mathbf{BiLSTM} (\bm{X}), \bm{O}=\mathbb{O}_{f}+\mathbb{O}_{b} \\
  \bm{c} &= tanh(\bm{O}) \times \bm{w}, y = \bm{O} \times \bm{c}
 \end{aligned}
\end{equation}

Attention is defined in Eq. \ref{atten}, where $\bm{X}$ is the input text features. $\mathbb{O}$ consists of $\mathbb{O}_{f}$ and $\mathbb{O}_{b}$ representing the forward and backward output of BiLSTM respectively. $\bm{w}$ is the learned weight vector from $\bm{O}$ and $c$ is the weighted context and $y$ is the final output with attention. 
\begin{table}[H]
  \caption{Parameter Settings of BiLSTM Model}
  \label{LSTM Params}
  \footnotesize
  \centering
  \begin{tabular}{cc}
    \toprule
    Layer Name & Parameter Settings \\
    \midrule
    BiLSTM & \tabincell{c}{Hidden: 128\\ Layers: 2 \\ Dropout: 0.5} \\
    \hline
     Attention   \\
    \hline
    Dropout & 0.5 \\
    \hline
    FC1 & \tabincell{c}{Out features: 128\\ Activation: ReLU}  \\
    \hline
    Dropout & 0.5 \\
    \hline
    FC2 & \tabincell{c}{Out features: 2\\ activation: ReLU} \\
    \bottomrule
  \end{tabular}
\end{table}

The detailed configuration of the proposed BiLSTM 
has been listed in Table \ref{LSTM Params}. 
The model consists of two BiLSTM layers, the output of which is fed into the attention layer for weight calculation.
The following two-layer FC network predicts whether the participant is in depression.

\subsection{Gate Recurrent Unit Neural Network}

A GRU model is utilized to process audio features.
The GRU model summarizes the audio embeddings to audio representations.  
Detailed configuration of the GRU model is shown in Table \ref{GRU Params}.
The proposed GRU model consists of two GRU layers, followed by a two-layer FC network that outputs binary labels predicting the presence of depression.

\begin{table}[H]
  \caption{Parameter Settings of GRU Model}
  \label{GRU Params}
  \footnotesize
  \centering
  \begin{tabular}{cc}
    \toprule
    Layer Name & Parameter Settings\\
    \midrule
    GRU & \tabincell{c}{Hidden: 256\\ Layers: 2 \\ Dropout: 0.5} \\
    \hline
    Dropout & 0.5 \\
    \hline
    FC1 & \tabincell{c}{Out features: 256\\ Activation: ReLU}  \\
    \hline
    Dropout & 0.5 \\
    \hline
    FC2 & \tabincell{c}{Out features: 2\\ activation: Softmax} \\
    \bottomrule
  \end{tabular}
  \vspace{-4mm}
\end{table}

\subsection{Multi-modal Fusion}
To integrate audio and text information
, representations generated by the last layer of GRU model and BiLSTM model are concatenated horizontally.
Modal attention is a weight vector trained 
to represent the importance of different modalities.
The dot product of attention vector and the concatenated representations produce the weighted representation, which is then passed to a one-layer FC network. 
Then, a loss function is derived 
as defined in Eq. \ref{loss}, where $m$ is the adopted modality, $\ell$ is the cross entropy loss function defined in Eq. \ref{ce}, $\bm{x}_m$ is the representation vectors of $m$, $\bm{\omega}_m$ is the weight of the FC network with respect to $m$ and $y$ is the ground-truth.
 
\begin{equation}
\label{loss}
\mathcal{L} = \sum_{m=\{audio, text\}} \ell_{ce} (\bm{x}_m, \bm{\omega}_m, y)
\end{equation}
\vspace{-1mm}
\begin{equation}
\label{ce}
\ell_{ce} = -\frac{1}{n} \sum_x [y \cdot logx + (1-y) \cdot log(1-x)]
\end{equation}

\section{Experiments and Results}
\label{experiments}

\subsection{DAIC-WoZ Datset}

DAIC-WoZ dataset is a public English depression dataset that contains recordings and transcripts of 142 participants, each of which is labeled with a PHQ-8 score \cite{Gratch2014}. 
PHQ-8 questionnaire is another popular questionnaire for depression screening but with less questions compared with SDS.
The participant whose PHQ-8 score greater than or equal to 10 is regarded in depression.
DAIC-WoZ dataset consists of a training set (30 depressed and 77 non-depressed), a development set (12 depressed and 23 non-depressed) and a test set which is not publicly available \cite{Gratch2014}.  
The experiments are finally performed on DAIC-WoZ and EATD-Corpus dataset, considering that AViD-Corpus doesn't provide text information.

\subsection{Data Imbalance}
\label{imbalance}
Data imbalance heavily exists in depression datasets.
Unbalanced datasets will introduce non-depressed preference to the trained classification models.
Therefore, the sizes of the depressed and non-depressed classes need to be balanced before training.
In this work, resampling is utilized to address the data imbalance issue.

For DAIC-WoZ dataset, samples in the two classes are equalized by group resampling.
Every 10 responses of one participant are grouped, along with the corresponding audios and text transcripts.
Samples are randomly selected from different groups of depressed participants without redundancy until the number of samples in the two classes are equivalent.
For example, a balanced training set consisting of 77 depressed samples and 77 non-depressed samples can be constructed from DAIC-WoZ dataset.
It should be noted that resampling is only performed on the training sets. 
In the testing phase, only one segment of audios and transcripts are randomly selected from each individual's responses in the development/test set and used for evaluation.

For EATD-Corpus, the method of rearranging volunteers' responses is adopted to increase the size of the depressed class.
The orders of three responses are rearranged and these rearranged responses are resampled to create new training samples.
Because there are 6 ways of response rearrangement for each individual, the size of the depressed class can be enlarged 6 times.

\subsection{Performance Evaluation on DAIC-WoZ Dataset}
In the text transcripts of DAIC-WoZ, 
responses to the same question are concatenated 
and encoded as the average of all three layer embeddings from ELMo \cite{matthew2018}.
A matrix of $N \times 1024$ is obtained for each participant, where $N$ is the number of questions.
To address the data imbalance issue, the matrix is divided into $m$ smaller matrices of size $10 \times 1024$, where $m$ is the integer of $N$ divided by $10$.
Resampling is performed on the divided matrices of depressed participants. 

Corresponding audio is segmented based on the timestamps in the text transcript.
NetVLAD is applied to generate 256-dimensional audio embeddings from extracted Mel spectrograms.
Similar to the text features, the matrix obtained for each participant is divided and resampling is performed. 

After extracting audio and text embeddings, a GRU model and a BiLSTM model with an attention layer are trained. 
Then, the 128-dimensional text and 256-dimensional audio representations 
are concatenated horizontally to 
train modal attention.
The dot products of the concatenated representations and modal attention are fed into the multi-modal network which produces binary labels. 

For performance comparison, F1 Score, Recall and Precision values are reported.
The performances of our approach together with some existing methods for depression detection are summarized in Table \ref{res1}.
From Table \ref{res1}, it can be seen that compared with the methods only adopting audio features, the proposed GRU model yields the highest performance with the F1 score equal to 0.77.
Compared with methods adopting only text features, the proposed BiLSTM model achieves the second-best performance with the F1 score equal to 0.83, which is merely 0.01 worse than the best method.
The proposed multi-modal fusion method produces the best result with its F1 score equal to 0.85.
Compared with the other method accepting both audio and text features, our method achieves a much better performance. In addition, the Recall values of our proposed single modality models and fusion model are close to 1. It indicates that our method can find out most of the depressed participants in practice.

\begin{table}[tbp]
\centering
\caption{Results of Experiments on DAIC-WoZ dataset}
\label{res1}
\resizebox{\linewidth}{!}{
\begin{tabular}{c|c|c|c|c}
\toprule
Features & Models & F1 Score & Recall & Precision  \\
\hline
\multirow{6}*{Audio} & Gaussian Staircase Model \cite{williamson2016} & 0.57 & - & - \\
~ & DepAudioNet  \cite{ma2016} & 0.52 & 1.00 & 0.35 \\
~ & Multi-modal LSTM  \cite{Al2018} & 0.63 & 0.56 & \textbf{0.71} \\
~ & SVM & 0.40 & 0.50 & 0.33 \\
~ & Decision Tree & 0.57 & 0.50 & 0.57\\
~ & Proposed GRU model & \textbf{0.77} & \textbf{1.00} & 0.63 \\
\hline
\multirow{6}*{Text} & Multi-modal LSTM \cite{Al2018} & 0.67 & 0.80 & 0.57 \\
~ & Cascade Random Forest  \cite{sun2017} & 0.55 & \textbf{0.89} & 0.40 \\
~ & Gaussian Staircase Model  \cite{williamson2016} & 0.84 & - & - \\
~ & SVM & 0.53 & 0.42 & 0.71 \\
~ & Decision Tree & 0.50 & 0.67 & 0.40 \\
~ & Proposed BiLSTM model & \textbf{0.83} & 0.83 & \textbf{0.83} \\
\hline
\multirow{2}*{Fusion} & Multi-modal LSTM \cite{Al2018} & 0.77 & 0.83 & 0.71 \\
~ & Proposed fusion model & \textbf{0.85} & \textbf{0.92} & \textbf{0.79} \\
\bottomrule
\end{tabular}
}
\vspace{-4mm}
\end{table}

\subsection{Performance Evaluation on EATD-Corpus Dataset}
The performances of the proposed method are further evaluated on EATD-Corpus with 3-fold cross validation.
The volunteers in the dataset are divided into three groups, two of which are used for training and the other one for testing.
As described in Section \ref{imbalance}, audios and transcripts of each depressed volunteer in the training set are rearranged and resampled.
Then, audio and text embeddings of size $3 \times 256$ and $3 \times 1024$ are extracted from the training set and test set.
The proposed GRU model and BiLSTM model are trained separately to generate representations, which are concatenated and passed to the multi-modal fusion network to output binary labels.

We implement the method introduced in \cite{Al2018} and evaluate its performance on EATD-Corpus for comparison.
The performances of three traditional classifiers, i.e. SVM, Random Forest, and Decision Tree, are also evaluated.
All these methods are evaluated using 3-fold cross validation.
The experimental results have been shown in Table \ref{res3}.
It can be seen that, when only using single modalities, the proposed GRU/BiLSTM model achieves the best performance compared with its counterparts.
When only audio features are considered, the F1 score of our method is 0.66, compared with the second best F1 score 0.50.
For text features, the F1 score of our method is 0.65, compared with the second best F1 score 0.64.
The results demonstrate the advantage of our method in dealing with depression detection problem.
Compared with the models using single modalities, our fusion model exhibits a much higher performance, with the F1 score increased to 0.71.
It is much higher than that of the method proposed in \cite{Al2018}. 
Similarly, the Recall values of the fusion model have also been significantly increased to 0.84, which indicates that our method can detect most depressive cases.
As a consequence, the fusion performance is only compared between two deep learning based methods.
These results demonstrate the effectiveness of the proposed fusion method.

The results from DAIC-WoZ and EATD-Corpus imply that our method has a powerful generalization ability and can be applied to different depression datasets.

\begin{table}[tbp]
\centering
\caption{Results of Experiments on EATD-Corpus}
\label{res3}
\resizebox{\linewidth}{!}{
\begin{tabular}{c|c|c|c|c}
\toprule
Features & Models & F1 Score & Recall & Precision  \\
\hline
\multirow{5}*{Audio} & Multi-modal LSTM \cite{Al2018} & 0.49 & 0.56 & 0.44 \\
~ & SVM & 0.46 & 0.41 & 0.54 \\
~ & RF & 0.50 & 0.53 & 0.48 \\
~ & Decision Tree & 0.45 & 0.44 & 0.47 \\
~ & Proposed GRU model & \textbf{0.66} & \textbf{0.78} & \textbf{0.57} \\
\hline
\multirow{5}*{Text} & Multi-modal LSTM \cite{Al2018} & 0.57	& 0.63 & 0.53 \\
~ & SVM & 0.64 & \textbf{1.00} & 0.48 \\
~ & RF & 0.57 & 0.53 & 0.61 \\
~ & Decision Tree & 0.49 & 0.43 & 0.59 \\
~ & Proposed BiLSTM model & \textbf{0.65} & 0.66 & \textbf{0.65} \\
\hline
\multirow{2}*{Fusion} & Multi-modal LSTM \cite{Al2018} & 0.57 & 0.67 & 0.49 \\
~ & Proposed fusion model & \textbf{0.71} & \textbf{0.84} & \textbf{0.62} \\
\bottomrule
\end{tabular}
}
\vspace{-4mm}
\end{table}

\section{Conclusion}
\label{sec:conclusion}

In this paper, we release the first public Chinese depression dataset EATD-Corpus.
It contains audio responses of 162 volunteers to three emotion related questions.
Text transcripts of audios are also extracted and manually corrected and supplied in EATD-Corpus.
Considering the rareness of the public multimedia depression datasets, EATD-Corpus provides valuable data for researchers in psychology and computer science who are engaged in depression study.

Beside, we propose a novel depression detection method which can detect depression state by analyzing audio signals and linguistic contents of the participants.
Our method simply encodes audio/text features into embeddings and doesn't rely on the contents of questions asked during the interview.
We evaluate the performance of the proposed method on two depression datasets, namely DAIC-WoZ and EATD-Corpus.
Experimental results demonstrate that the proposed method is of great effectiveness.
In the future, we intend to build an APP that allows users to self-detect their depressive states based on the proposed method.

\bibliographystyle{IEEEbib}
\bibliography{strings,refs}

\end{document}